\documentclass[journal=jacsat,manuscript=article]{achemso}
\usepackage{graphicx}
\usepackage{dcolumn}
\usepackage{bm}
\usepackage[table]{xcolor}
\usepackage{color, colortbl}
\usepackage[utf8]{inputenc}
\usepackage{gensymb}
\usepackage{xcolor}
\usepackage{ulem}
\usepackage{makecell}
\usepackage{tabularx,colortbl}
\usepackage{blindtext}
\usepackage{booktabs}
\usepackage{chemformula} 
\usepackage[T1]{fontenc}

\title{Static and dynamic magnetization control of extrinsic multiferroics by the converse magneto-photostrictive effect.}

\author{Matthieu Liparo}
\affiliation{Univ. Brest, Laboratoire d'Optique et de Magnétisme (OPTIMAG), UR 938, 29200 Brest, France}
\alsoaffiliation{Cr Research Group, Department of Physics, University of Johannesburg, PO Box 524, Auckland Park 2006, South Africa}
 
    \author{Jean-Philippe Jay}
    \author{Matthieu Dubreuil}
     \affiliation{Univ. Brest, Laboratoire d'Optique et de Magnétisme (OPTIMAG), UR 938, 29200 Brest, France}
    \author{Ga\"elle Simon}
     \affiliation{Univ. Brest, Service général des plateformes technologiques, Service RMN-RPE, 29200 Brest, France}
    \author{Alain Fessant}
\author{Walaa Jahjah}
\author{Yann Le Grand}
    \affiliation{Univ. Brest, Laboratoire d'Optique et de Magnétisme (OPTIMAG), UR 938, 29200 Brest, France}
\author{Charles Sheppard}
\author{Aletta R. E. Prinsloo}
    \affiliation{Cr Research Group, Department of Physics, University of Johannesburg, PO Box 524, Auckland Park 2006, South Africa}
    \author{Vincent Vlaminck}
    \author{Vincent Castel}
    \author{Loic Temdie-Kom}
    \author{Guillaume Bourcin}
\affiliation{IMT-Atlantique, Campus de Brest, Département Micro-Ondes, Technopole Brest-Iroise, CS83818, 29238 Brest Cedex 03, France}
      \author{David Spenato}
  \affiliation{Univ. Brest, Laboratoire d'Optique et de Magnétisme (OPTIMAG), UR 938, 29200 Brest, France}
\author{David T. Dekadjevi}

    \affiliation{Univ. Brest, Laboratoire d'Optique et de Magnétisme (OPTIMAG), UR 938, 29200 Brest, France}
    \alsoaffiliation
    {Cr Research Group, Department of Physics, University of Johannesburg, PO Box 524, Auckland Park 2006, South Africa}
    \email{david.dekadjevi@univ-brest.fr}

\begin{document}

\date{\today}

\newpage

\begin{abstract}

In this work, photostrictive manipulations of static and dynamic magnetic properties are demonstrated in an extrinsic multiferroic composite. The photostriction is achieved with visible light in the blue region of the spectrum. The composites consist of 5 nm or 10 nm magnetostrictive Fe$_{81}$Ga$_{19}$ thin films coupled to a piezoelectric (011)-Pb(Mg$_{1/3}$Nb$_{2/3}$)O$_3$-Pb(Zr,Ti)O$_3$ substrate. Angular dependent magnetization reversals properties are largely enhanced or reduced under a converse magneto-photostrictive effect (CMPE). The CMPE strength is analysed with a novel coefficient named the converse magneto-photostrictive coupling coefficient. This coefficient is proposed as a general approach to analyse and to compare different extrinsic multiferroics under the converse magneto-photostrictive effect. Its thickness dependence reveals that the CMPE strength decreases with an increase of the Fe$_{81}$Ga$_{19}$ thickness. Experimental evidence for a control of dynamic magnetic properties under CMPE is then revealed by ferromagnetic resonance measurements. Resonant fields are shifted under CMPE, whereas their linewidths remain constant. Furthermore, resonant field shifts can be either positive or negative depending on the in-plane angle. The largest shift under CMPE of +5.7~\% is obtained for the 5~nm sample. Our study shows that the CMPE provides an efficient approach for a control of not only the static but also the dynamic magnetic properties in extrinsic multiferroics.

\end{abstract}

\newpage

\maketitle

The "digital world" has a large environmental cost and the predictions shows that it will be responsible for more than 20 \%  of the world's electricity consumption by 2030~\cite{challe6010117, LANGE2020106760}. A paradigm to achieve energy-efficient information processing, computation, communication or signal generation is envisaged by using magnetostrictive nanomagnets, and manipulating their magnetic anisotropy through strain~\cite{Bukharaev_Review_IOP2018, D_Souza_Nanotechnology_Review_2018, Bandyopadhyay_APL_2021, Trassin_Review_MultiF_PhysSciRev_2021}.
The strain control of magnetic anisotropy constitutes a keystone of the research area named magnetic straintronics, where strain-induced physical effects in solids are used to develop next-generation devices for energy-saving technologies including energy efficient information storage and sensor. Many studies on magnetic straintronics for energy saving involve a family of materials named extrinsic multiferroics (ExMF), and the research concerning this family has attracted a lot of interest in the last few decades~\cite{Spaldin05, Fusil2014, fiebig16, Spaldin_Review_2020}. 

ExMF simultaneously display two or more ferroic orders (\textit{eg.} ferroelectric, ferromagnetic or/and ferroelastic), and their magnetic properties can be modified through strain~\cite{eerenstein06}.
These are multi-functional materials with magnetostrictive and piezoelectric phases. The mechanism of strain driven magnetic properties in ExMF relies on the strain being transferred to the magnetostrictive phase and, in turn, induces inverse magnetostriction (the Villari effect \cite{Lacheisserie1993}), which translates into a change in magnetic properties. Thus, understanding and controlling the magnetic properties through strain in these multiferroics are of great interest for energy efficient applied issues~\cite{Thiele2007, Yang2009, Wu2011, Zhang2014, Nan2014, alberca2015, Yang2015, Staruch2016, Iurchuk_PRL2016, Biswas2017, Bukharaev_Review_IOP2018, D_Souza_Nanotechnology_Review_2018, Zhang_APL2020_PMNPT-Ni, Hu_ReviewMF_2019, Bandyopadhyay_APL_2021, Trassin_Review_MultiF_PhysSciRev_2021}. Strain can be introduced in the ExMF using a variety of stimuli, however, it is important that the strain is produced using an energy-efficient method in order to reduce environmental costs.\

Until now, a large number of studies focus on the converse magneto-electric effect (CME), that is the magnetization control in strain-mediated composites with an electrical stimulus~\cite{Thiele2007, Yang2009, Wu2011, Zhang2014, Nan2014, alberca2015, Yang2015, Staruch2016, Biswas2017, Bukharaev_Review_IOP2018, D_Souza_Nanotechnology_Review_2018, Bandyopadhyay_APL_2021, Trassin_Review_MultiF_PhysSciRev_2021}. The magneto-electric coupling occurs when an applied electric field induces strain in the piezoelectric phase through the piezoelectric effect. 
In order to quantify the relative magnetization change upon applying an electric field, a converse magneto-electric coupling coefficient can be expressed as 
$\alpha_{CME}=\mu_0 \partial{M} / \partial{E} $ 
(in s.m$^{-1}$), 
with $\mu_{0}$ the vacuum permeability, $M$ the magnetization and $E$ the electric field~\cite{Eerenstein2007, Thiele2007,  alberca2015, Staruch2016, Jahjah2020_PRA}. During the past few years, more and more research on ExMF $\alpha_{CME}$ were performed~\cite{Eerenstein2007, Thiele2007, Yang2009, heron2014, cherifi2014, alberca2015, Zhang2014, Staruch2016, wu2016, wei2016, Lian2018-2, Wang2019, Jahjah2020_PRA}. A major objective of these studies was to understand driving mechanisms of the magneto-electric coupling phenomenon and to compare its efficiency in different materials. In general, previous studies have shown that electric field control of static and dynamic magnetic properties through strain is of interest in the context of energy saving technologies~\cite{Heron2021_reviewMFenergy}. However, large electric fields are required for this control and different mechanisms for strain control of magnetic properties would be of prime interest.\ 
 
 In ExMF, only a few studies focused on the optical control of ferromagnetic magnetization using the photostrictive effect and the understanding thereof~\cite{Zhang_APL2020_PMNPT-Ni, Iurchuk_PRL2016}. The photostrictive effect is the light-induced non-thermal dimension change of materials. The mechanism responsible for the phenomenon is different depending on the material. In ferroelectric materials, photostriction arises from a combination of photovoltaic and reverse piezoelectric effects and it is present in ExMF ferroelectric substrates~\cite{Kundys_Review_Photostrictive_2015, Chen_Review_Photostrictive2021}. Response times of photostrictive effects in ferroelectrics are rather slow, from tenths of a second to minutes. This limitation originates from the slow building up of the photovoltage across the material in order to generate the strain. It was recently demonstrated that this limitation can be overcome by utilizing the faster microscale photovoltaic response and construction of local photostrictive strain~\cite{Liew_Small_2022}.\


 
  
 The change of magnetic properties under the photostrictive effect occurs when light induces strain in the piezoelectric phase through the photostrictive effect. The dimensional changes are transferred to the magnetostrictive phase and, in turn, induces inverse magnetostriction, which translates into a change in magnetic properties. The change of magnetic properties under the photostrictive effect defines the converse magneto-photostrictive effect (CMPE), which can be schematically written as (light$/$mechanical) $\times$ (mechanical$/$magnetic). Two previous studies have shown that converse magneto-photostrictive effect does occur in ExMF~\cite{Zhang_APL2020_PMNPT-Ni, Iurchuk_PRL2016}. These studies revealed modifications of the thin films ferromagnetic magnetization reversal loops, under substrate illumination in Ni(11~nm)/BiFeO$_3$~\cite{Iurchuk_PRL2016} and in Ni(11~nm)/PMN-PT ~\cite{Zhang_APL2020_PMNPT-Ni} samples. However, CMPE key properties remain largely unexplored in ExMF. In particular, CMPE anisotropic properties in ExMF have not yet been studied, despite the importance of anisotropic properties in all (magnetic) systems. Furthermore, to encompass and describe the CMPE, a converse magneto-photostrictive coefficient would be relevant, such as $\alpha_{CME}$ is relevant for magneto-electric phenomena. A converse magneto-photostrictive coefficient has not yet been defined, nor studied. Furthermore, the control of dynamic properties using the CMPE has never been demonstrated. Such a control would provide an alternative to the electric field control of magnetic properties, and would pave a way to alternative technologies and further energy savings in information technologies.\ 

In this manuscript, we report on the angular dependent photostrictive manipulations of static and dynamic magnetic properties under visible light (in the blue region of the spectrum) in ExMF, which consist of magnetostrictive Fe$_{81}$Ga$_{19}$ (FeGa) thin films of 5~nm and 10~nm prepared on a (011)-Pb(Mg$_{1/3}$Nb$_{2/3}$)O$_3$-Pb(Zr,Ti)O$_3$ (PMN-PZT) piezoelectric substrate. For the 5~nm and 10~nm FeGa thicknesses, the evolution of anisotropic properties is presented through the angular dependencies of the magnetization reversal (MRev) properties in the dark state, that is without illumination, and under a laser illumination. Then, a converse magneto-photostrictive (CMP) coupling coefficient, $\alpha_{CMP}^\lambda$, is proposed. In order to determine its anisotropic characteristics and thickness dependence, angular dependencies, signs and magnitudes of $\alpha_{CMP}^\lambda$ are probed for the 5~nm and 10~nm FeGa thicknesses. Finally, for the 5~nm thick FeGa, a comparison between magneto-electric and magneto-photostrictive effects is made through the analysis of the angular dependencies of $\alpha_{CMP}^\lambda$ and $\alpha_{CME}$. The thickness dependence of the maximum value of the $\alpha_{CMP}^\lambda$ is established. Finally, using ferromagnetic resonance measurements, experimental evidence for the control of the dynamic magnetic properties by CMPE is revealed.\

In this study, a rhombohedral PMN-PZT single crystal is chosen as a substrate since it is a relaxor-PbTiO$_3$ (relaxor-PT) based ferroelectric. It is used for its excellent piezoelectric properties \cite{Zhang2007,Zhang2010}. In particular, (011)  PMN-PZT demonstrates a large in-plane anisotropic piezostrain~\cite{Zhang2010, Shanthi2008}. For the magnetostrictive phase, 5~nm and 10~nm FeGa thin films were chosen as these demonstrate remarkable properties such as low hysteresis, large magnetostriction and good tensile strength \cite{Clark2000,Atulasimha2011,jay2019,Dekadjevi2021_PRA}. In addition, the FeGa/PMN-PZT heterostructure studied in this manuscript was shown to exhibit a large magneto-electric coupling coefficient~\cite{Jahjah2020_PRA}. Consequently, if an optical stimulus would provide non-negligible strain through the PMN-PZT photostriction, it is likely to lead to a non-negligible control of FeGa magnetic properties through the inverse magnetostrictive effects. The samples were prepared by depositing the magnetostrictive FeGa thin films onto the PMN-PZT substrates, using radio-frequency magnetron sputtering. The growth was carried out under an in-plane magnetic field $H$\textsubscript{dep} = 2.4 kA.m$^{-1}$ along the [100] direction of the PMN-PZT substrate. In the rest of this manuscript, $\varphi$ is the angle between the applied magnetic field $H$ and the [100] direction of the PMN-PZT substrate. Static magnetic measurements at room temperature were done using the magneto-optic Kerr effect (MOKE). The CMPE control of static magnetic properties was studied by illuminating the samples with an intensity $I_1$ of 0.6~W.cm$^{-2}$ from a 410~nm laser diode. Dynamic measurements were obtained using an electron paramagnetic resonance spectrometer operating at X-band (9.3 GHz). For dynamic measurements, the CMPE was studied by illuminating the samples with the laser diode (the one previously described in the Static magnetic measurements) and it was also studied by illuminating the samples with a LED. The LED wavelength is 405~nm and the intensity of the LED illumination is 7 W.cm$^{-2}$. MOKE measurements could not be performed with the LED due to space restrictions in the experimental set-up.\

\newpage

\section{Results}

\subsection{Magnetization reversals in the dark state and under illumination.}

\begin{figure*} [ht]
\centering\includegraphics[width=16.5cm]{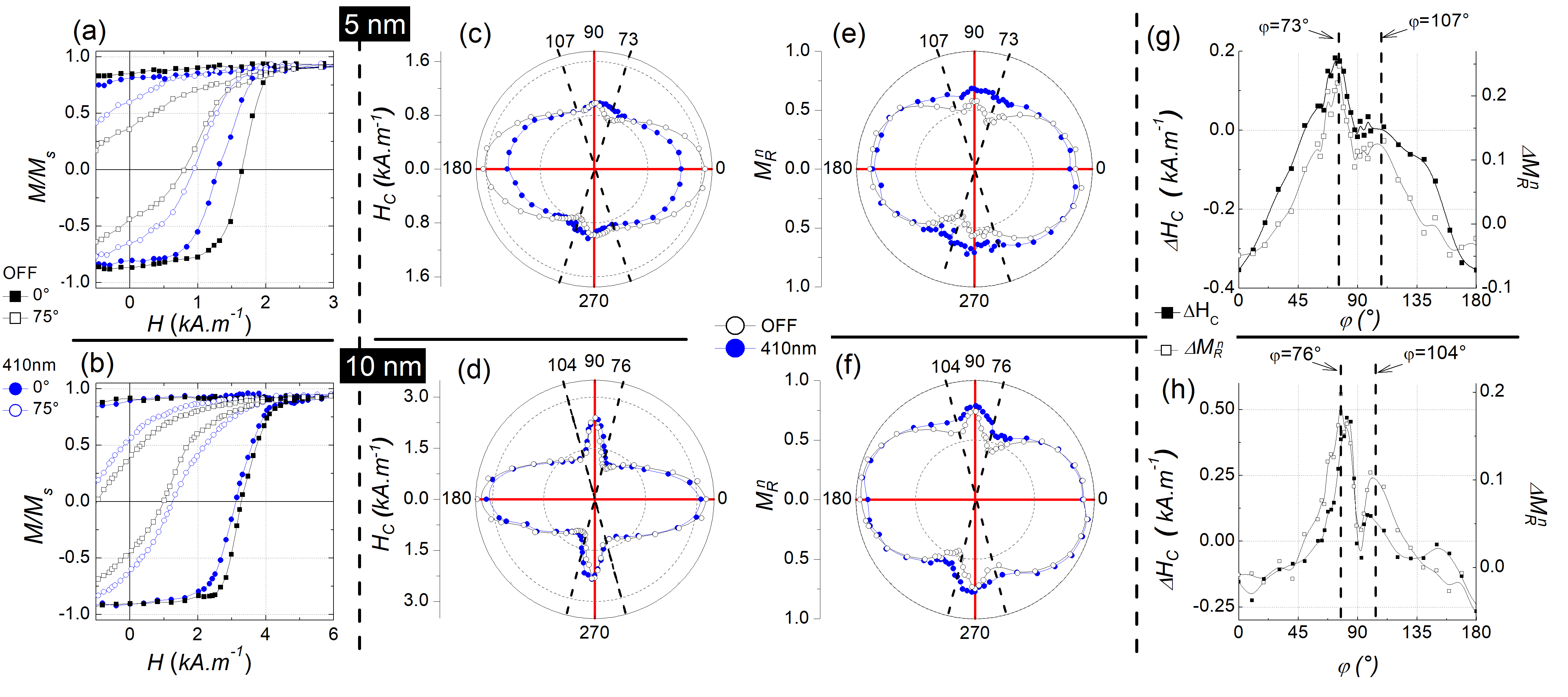}
\caption{\label{fig:fig1}(a) and (b) Zoomed hysteresis loops of the normalized magnetization reversal of the Ta(10 nm)/FeGa(5 nm and 10 nm)/PMN-PZT(0.3 mm), measured in-plane with the magnetic field $H$, respectively, parallel to [100] ($\varphi$ = 0\degree) and $\varphi$ = 75\degree, under CMPE (i.e. under illumination at 410~nm) and in the dark state (OFF). (For information: the entire M-H loops are presented in Figure S2 of the supplementary materials). (c) and (d) $H_c$ polar plots for both samples, under CMPE and in the dark state. (e) and (f) $M_R^n$ polar plot for both samples, under CMPE and in the dark state. (g) and (h) $\Delta H_c$ and $\Delta M_R^n$ for both samples, under CMPE and in the dark state.}
\end{figure*}

Figures~\ref{fig:fig1} (a) and (b) show MRev loops in the dark state (i.e. light "OFF"), under the illumination with the 410~nm laser diode, for both FeGa thicknesses and at two different angles $\varphi$. It can be observed from Figures~\ref{fig:fig1} that the dark state MRev, in $\varphi$ = 0$\degree$ and 75$\degree$, for the 5~nm and 10~nm samples are modified under laser illumination. These modifications did not evolve with time (see Figure S3 of the supplemental materials for more information). Also, the maximum increase in sample temperature due to the illumination was (0.6 $\pm$ 0.1)K during measurements (see Figure S4 of the supplemental materials for more information). It should be noted that we previously showed that such a limited temperature increase does not significantly modify the magnetization reversal of FeGa thin films grown on PMN-PZT substrates~\cite{Jahjah2020_PRA}. The light-induced changes shown in Figures~\ref{fig:fig1} (a) and (b) reveal a non-thermal and stable control of the FeGa magnetic properties through the PMN-PZT photostriction. In this manuscript, the modification of the ExMF magnetic properties under photostriction is named the converse magneto-photostrictive effect (CMPE), in the spirit of the well-known converse magneto-electric effect previously reported in ExMF~\cite{Thiele2007, Yang2009, Wu2011, Zhang2014, Nan2014, alberca2015, Yang2015, Staruch2016, Biswas2017, Bukharaev_Review_IOP2018, D_Souza_Nanotechnology_Review_2018, Bandyopadhyay_APL_2021}.\

Under CMPE, both samples reveal a decrease in the M-H loop area for magnetic fields applied along $\varphi=0\degree$ but an increase in the M-H loop area along $\varphi=75\degree$. It may be noted here that such an angular phenomenon due to illumination has not previously been reported in ExMF. A previous study on a Ni/PMN-PT heterostructure probed the light-induced changes of MRev along two different axes but did not report such a MRev angular-dependent property~\cite{Zhang_APL2020_PMNPT-Ni}. Here, the observed FeGa/PMN-PZT CMPE is angular-dependent not only in magnitude but also ``in sign'' (in the sense of a decrease or increase in MRev characteristic properties). For both samples, the CMPE does not affect the normalized remanent magnetization ($M_R^n$) along $\varphi=0\degree$ but increases $M_R^n$ along $\varphi=75\degree$. For both samples, the CMPE reduces the coercive field ($H_c$) along $\varphi=0\degree$ but increases $H_c$ along $\varphi=75\degree$.\

In order to further understand this angular-dependent CMPE, MRev angular dependencies were probed for both samples. In the dark state, and under CMPE, $H_c$ and $M_R^n$ values have been obtained from each M-H loop as shown in Figure~\ref{fig:fig1} (c)-(f). Probing these $H_c$ and $M_R^n$ angular dependencies is of interest because they are related to magnetic anisotropic properties. In previous studies on FeGa thin films, it was shown that FeGa anisotropy configurations involve an uniaxial anisotropy (UA), a cubic anisotropy (CA) and a random anisotropy (RA) \cite{jay2019, Jahjah2020_PRA, Dekadjevi2021_PRA}. An UA was found to be driven by $H$\textsubscript{dep}. The UA results in the presence of $H_c$ and $M_R^n$ maxima (minima) along a single axis named the easy axis (the hard axis). The CA was found to originate from a (110) FeGa preferred orientation. The CA results in the presence of $H_c$ and $M_R^n$ maxima (minima) along two distinct easy (hard) axes. The RA arises from the polycristalline nature as the FeGa (110) orientation is only preferential. The RA does not result in $H_c$ and $M_R^n$ extrema. FeGa thin films crystallographic textures have been shown to be thickness-dependent, and hence their anisotropic configurations too~\cite{jay2019}. In the study presented here, probing $H_c$ and $M_R^n$ angular dependencies under CMPE for two different FeGa thicknesses should provide a way to assess FeGa anisotropy modifications under CMPE.\

Here, for both thicknesses, $H_c$ and $M_R^n$ angular dependencies exhibit global (local) maxima lying along the $0\degree$ ($90\degree$) axis corresponding to the global (local) easy axis, as shown in Figure~\ref{fig:fig1}. The presence of these maxima of different magnitudes reveals the presence of UA and CA \cite{Ryon_JAP2019,Dekadjevi2021_PRA}.
For both thicknesses, the presence of two hard axes on each side of the easy axes is also shown by the angular-dependent study. Indeed, $H_c$ and $M_R^n$ angular dependencies of the 5~nm (10~nm) sample exhibit global minima along the $73\degree$ ($76\degree$) axis and local minima along the  $107\degree$ ($104\degree$) axis. The global (local) minima are along the global (local) hard axis.

Beyond these common points for both thicknesses which concern H$_c$ and $M_R^n$ extrema, all dark state shapes in Figure~\ref{fig:fig1} are thickness-dependent. These different shapes result from different anisotropy configurations involving UA, CA and RA anisotropy components. Different anisotropy configurations are related to structural thickness dependencies of the sputtered FeGa thin films discussed in the previous paragraph.\

CMPE brings about significant modifications to the magnetic properties. For both thicknesses, all angular dependencies are re-shaped by the illumination showing changes of the FeGa/PMN-PZT dark state anisotropy under CMPE. Along global easy axes and for both thicknesses, large decreases of H$_c$ are observed and $M_R^n$ values remain quasi-constant. However, along global hard axes and for both thicknesses, large increases of $H_c$ and $M_R^n$ are present. The largest ${H_c}$ modification under CMPE, $\Delta{H_c}$, is an increase of +0.46 kA.m$^{-1}$ (i.e. a 37\% relative increase). It is obtained along the 10~nm global hard axis as shown in Figure \ref{fig:fig1} (g) and (h). The largest $M_R^n$ modification under CMPE, $\Delta{M_R^n}$, is an increase of +0.25 (i.e. a 60\% relative increase). It is obtained along the 5~nm global hard axis. For both thicknesses, $\Delta{H_c}$ and $\Delta{M_R^n}$ negative values are observed with their minima always on the global easy axes. The analysis of Figure~\ref{fig:fig1} indicates that the ExMF FeGa/PMN-PZT studied here shows significant angular-dependent changes in magnetization reversals under CMPE, with H$_c$ and $M_R^n$ extrema modifications along easy and hard axes.\

\newpage
\subsection{Converse magneto-photostrictive coupling coefficient : definition}

\begin{figure}
\centering\includegraphics[width=12cm]{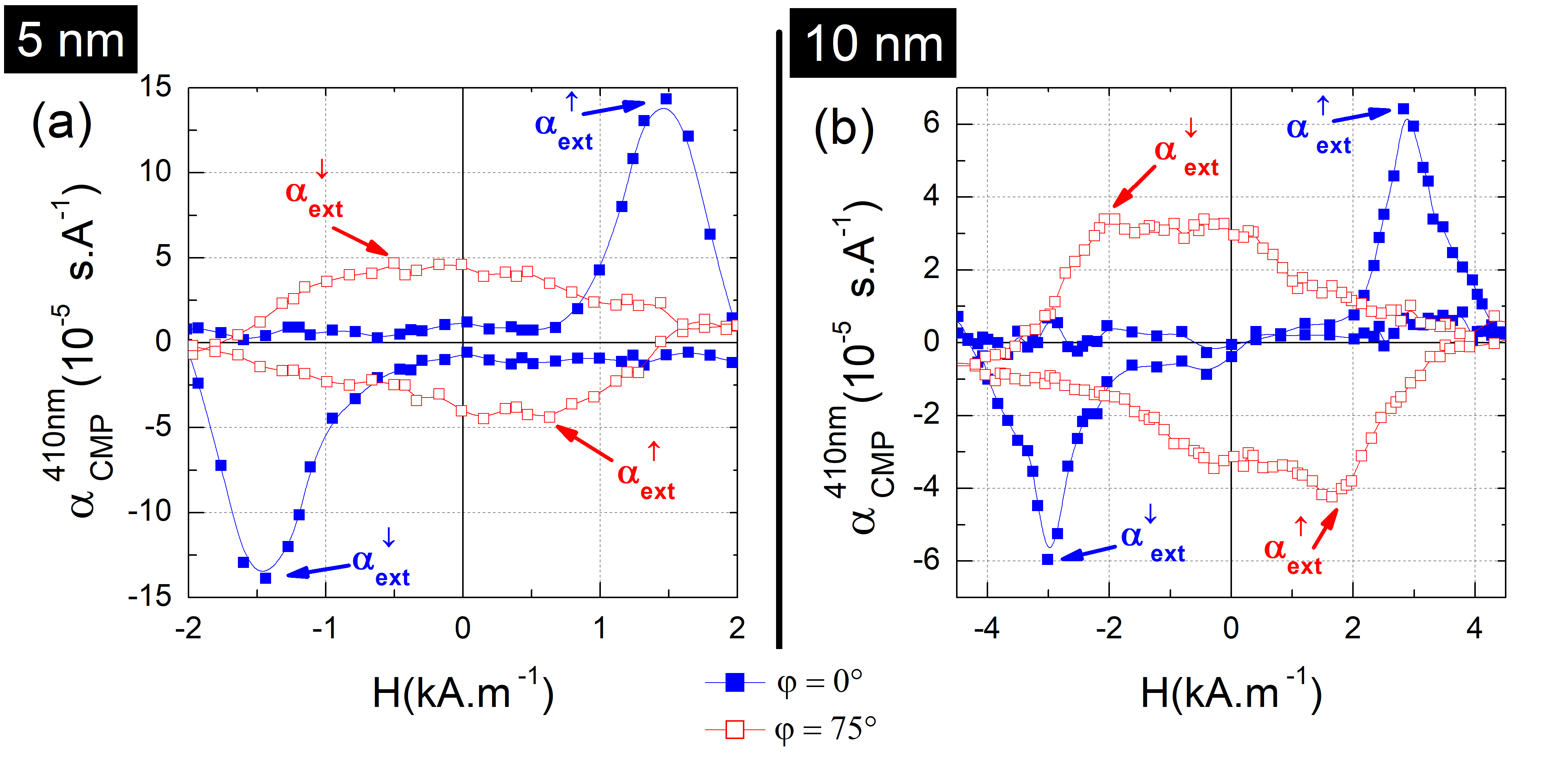}
\caption{\label{fig:fig2} The converse magneto-photostrictive coupling coefficient $\alpha_{CMP}^{410~nm}$ as estimated from $M(H)$ data in Figure~\ref{fig:fig1} for the Ta(10 nm)/FeGa(5 nm (a); 10 nm (b))/PMN-PZT samples.}
\end{figure}

As previously introduced, the converse magnetoelectric coupling coefficient $\alpha_{CME}$ (expressed in s.m$^{-1}$) was defined to quantify the electric-field-induced variation in the magnetic properties~\cite{Fiebig_2005,Chen_Review_Photostrictive2021}. This coefficient was calculated as $\alpha_{CME}(H)=\mu_0 \Delta{M(H)} / \Delta{E} $, with $\Delta{M}(H)$, the change in magnetization at a field $H$ under a change of electric field $\Delta{E}$ \cite{Eerenstein2007, Thiele2007, alberca2015, Staruch2016, Jahjah2020_PRA}. Thus, it represents the variation of the magnetization under an applied electric field. This coefficient allows an evaluation of the electric stimulus efficiency, provides a way to correlate this efficiency with materials fundamental properties, and provides a convenient approach for comparing different materials. It has proven to be interesting and is now widely used for ExMF. In a similar way, we propose here to assess the light-induced variation of the magnetic properties through photostriction with a converse magneto-photostrictive coupling coefficient, $\alpha_{CMP}^\lambda=\mu_0 \partial{M} / \partial{I}$ (expressed in s.A$^{-1}$) with $I$ the light intensity.

The  magneto-photostrictive coefficient $\alpha_{CMP}^\lambda$ is calculated by~:
\begin{equation}
\label{alpha_equation}
\alpha_{CMP}^{\lambda}(H)= \mu_0\frac{\Delta{M(H)}}{\Delta{I}}= \mu_0\frac{M_{I=I_1}(H)-M_{I=I_0}(H)}{I_1-I_0}
\end{equation}
with $\Delta{M(H)}$, the change in magnetization at a field $H$ under a change of  light intensity $\Delta{I}$ at a wavelength $\lambda$. In our experimental work, $\alpha_{CMP}^{410}$(H) can be determined with $I_1=0.6$~W.cm$^{-2}$ and $I_0=0$. The relative change in magnetization is directly computed from the measured MRev loops in the dark state and under CMPE. FeGa thin films studied here have a saturation magnetization $\mu_0 M\textsubscript{s}=1.15$~T~\cite{jay2019}. Thus, the $\alpha_{CMP}^{410~nm}$ values can be calculated.\

Figure~\ref{fig:fig2} shows $\alpha_{CMP}^{410~nm}$(H) at 0$\degree$ and 75$\degree$ for the two FeGa thicknesses. These were determined using the M-H loops shown in Figure~\ref{fig:fig1}. Significant dependencies of $\alpha_{CMP}^{410~nm}$ on the external magnetic field H are observed. For both thicknesses and for both angles, $\alpha_{CMP}^{410~nm}(H)$ exhibits two extrema of opposite signs. The first (second) extremum, $\alpha_{ext}^{\downarrow}$ ($\alpha_{ext}^{\uparrow}$), occurs when $H$ decreases (increases) as shown in Figure~\ref{fig:fig2} (a) and (b). Over the full angular range, $\alpha_{ext}^{\downarrow}$($\varphi)$ and $\alpha_{ext}^{\uparrow}$($\varphi)$ are of the same magnitude within experimental uncertainty but of opposite sign as shown in Figure~\ref{fig:fig3} (a) and (b). This is a consequence of M-H loops symmetry. $\alpha_{ext}^{\downarrow}$ and $\alpha_{ext}^{\uparrow}$ angular dependencies exhibit extrema along easy and hard axes. The extrema magnitudes are thickness-dependent.\

\newpage
\subsection{Angular dependence of converse magneto-photostrictive coupling coefficient maxima.}

\begin{figure} [ht]
\centering\includegraphics[width=10cm]{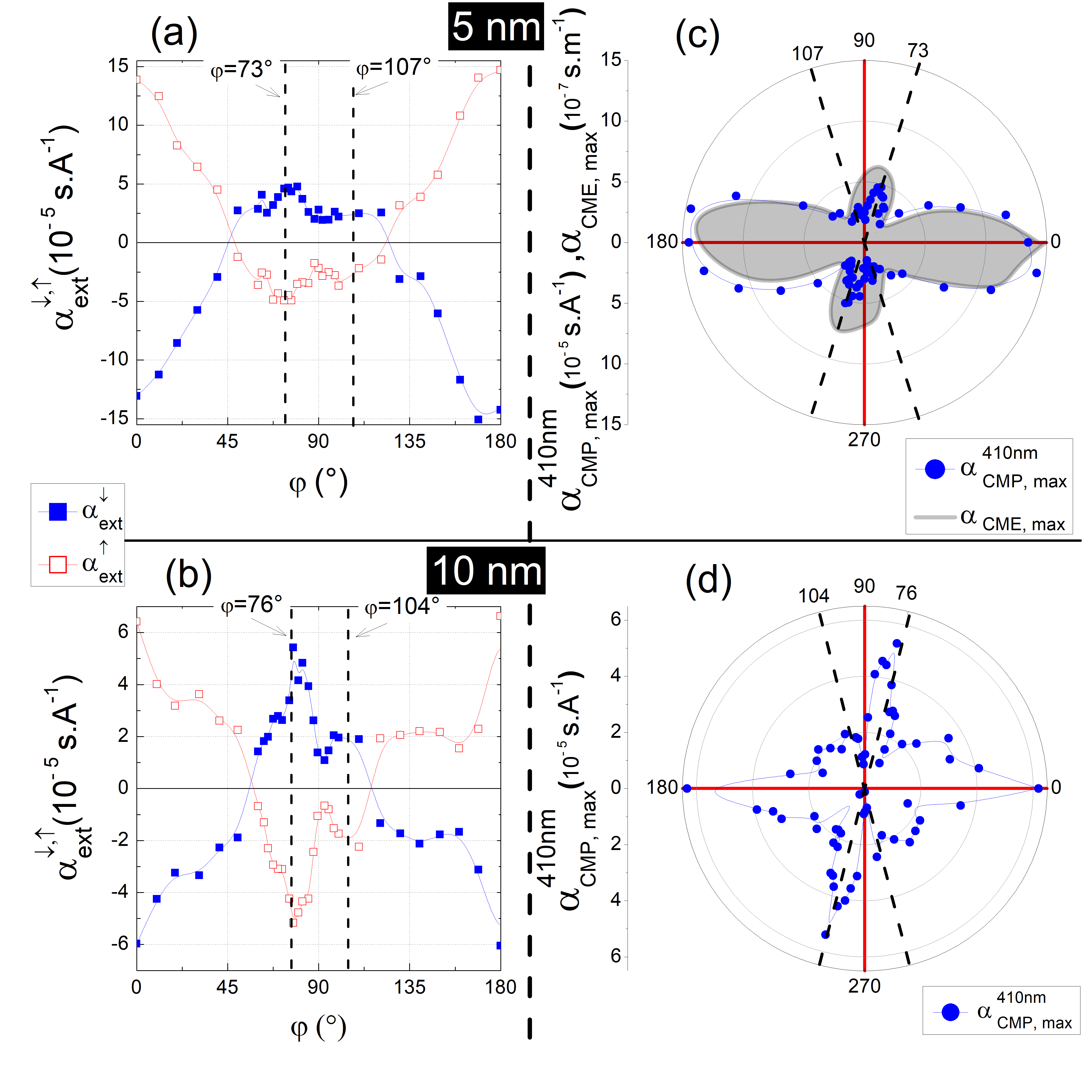}
\caption{\label{fig:fig3} Angular dependencies of $\alpha_{ext}^{\downarrow}$ and  $\alpha_{ext}^{\uparrow}$ for the (a) 5~nm and (b) 10~nm samples. Angular dependencies of $\alpha^{410~nm}_{CMP, max}$ for the (c) 5~nm and (d) 10~nm samples. (c) The gray shaded area is an eye guide to indicate the $\alpha_{CME, max}$ values previously reported for a 5~nm FeGa sample (from~\cite{Jahjah2020_PRA}).}
\end{figure}

{In order to further understand the CMPE through the use of $\alpha_{CMP}^{410~nm}$, let us first define the maximum value of the $\alpha_{CMP}^{410~nm}$(H) for a given angle $\varphi$ as : $\alpha^{410~nm}_{CMP, max}=(\lvert \alpha_{ext}^{\downarrow}\rvert + \lvert\alpha_{ext}^{\uparrow}\rvert)/2$. The magnitude of the $\alpha^{410~nm}_{CMP, max}$ angular dependence is proportional to the angular-dependent maximum change in magnetization inferred by the illumination as stated by Equation~\ref{alpha_equation} (as long as the illumination is not angular-dependent). For both thicknesses, $\alpha^{410~nm}_{CMP, max}$ extrema values are consistently present in the vicinity of the global easy and hard axes as shown in Figure~\ref{fig:fig3} (c) and (d). Maximum values are obtained in the vicinity of global easy axes.}

\begin{table} [ht]

\centering

\setlength\extrarowheight{0pt}

\begin{tabular}{|c|c|c|c|}

\cline{2-4} \multicolumn{1}{c|}{} & Units & FeGa (5nm) & FeGa (10nm)\\

\hline

Maximum of $\alpha_{CMP,max}^{410\ nm} $ & $10^{-5}$ s. A$^{-1}$ & $14.4 \pm 0.1$ & $6.3 \pm 0.1$\\

\hline

Maximum of $\mu_0 \Delta M$ & T & $0.87 \pm 0.02$ & $0.38\pm 0.02$\\

\hline

Maximum of $\frac{\Delta M}{M}$ & \% & $76 \pm 3$  & $33 \pm 2$\\

\hline

\end{tabular}

\caption{\label{alphavalue}Thickness dependence of the largest converse magneto-photostrictive coupling coefficient maxima (Maximum of $\alpha^{410~nm}_{CMP, max}$), of the largest change of magnetization change under CMPE (Maximum of $\mu_0 \Delta{M}$) and of the largest relative change of magnetization under CMPE (Maximum of $\frac{\Delta{M}}{{M}}$).}

\end{table}

\vspace{2cm}


{$\alpha^{410~nm}_{CMP, max}$ angular dependencies are thickness-dependent not only in shapes but also in magnitudes as shown in Figure~\ref{fig:fig3} and in Table~\ref{alphavalue}. It can be understood since the FeGa thin films magnetostrictive coefficient was found to decrease when the thickness increases~\cite{jay2019} and a thickness-dependent dark state FeGa anisotropy configuration is present. About magnitudes, considering that an Fega thin film magnetization of 1.15~T is fully reversed due to an illumination of 0.6~W.cm$^{-2}$, the greatest $\alpha_{CMP}^{410~nm}$ for FeGa thin films can be found from Equation~\ref{alpha_equation} as  38$\times$10$^{-5}$~s.A$^{-1}$. Relatively to this potential value, the greatest $\alpha^{410~nm}_{CMP, max}$ obtained here are significant, indicating the CMPE efficiency for the ExMF examined in our study. This efficiency is also indicated by the maximum change of magnetization as shown in Table 1.}\


{With respect to efficiency, for an ExMF it would be of interest to compare the angular-dependent maximum change in magnetization inferred by different stimuli. In deed, as long as both stimuli are kept constant, $\alpha_{CMP}^\lambda$ and $\alpha_{CME}$ angular-dependent shapes are respectively equal to the shapes of angular-dependent maximum change in magnetization inferred by both stimuli. The $\alpha_{CME}$ angular dependence under an electric field (E) of 6.5~kV.cm$^{-1}$ in the dark state and for the 5~nm sample was previously reported~\cite{Jahjah2020_PRA}. It is shown in Figure~\ref{fig:fig3} (c). Its shape is similar to that of the $\alpha^{410~nm}_{CMP, max}$ angular dependence shape. Such a similarity indicates comparable anisotropy modifications at stake under CMPE and CME. To compare the efficiency of both stimuli, the change of magnetization under CME and CMPE are given : a maximum (relative) change of magnetization of 0.87 $\pm$ 0.02~T (76\%) is obtained under CMPE, and a maximum (relative) change of magnetization of 0.94 $\pm$ 0.02~T (82\%) is obtained under CME. These values are an indication of the efficiency for both converse effects (i.e. both stimuli).}





\newpage

\subsection{Experimental evidence for a control of the ferromagnetic resonance field by CMPE.}

\begin{figure}
\centering \includegraphics[width=7cm]{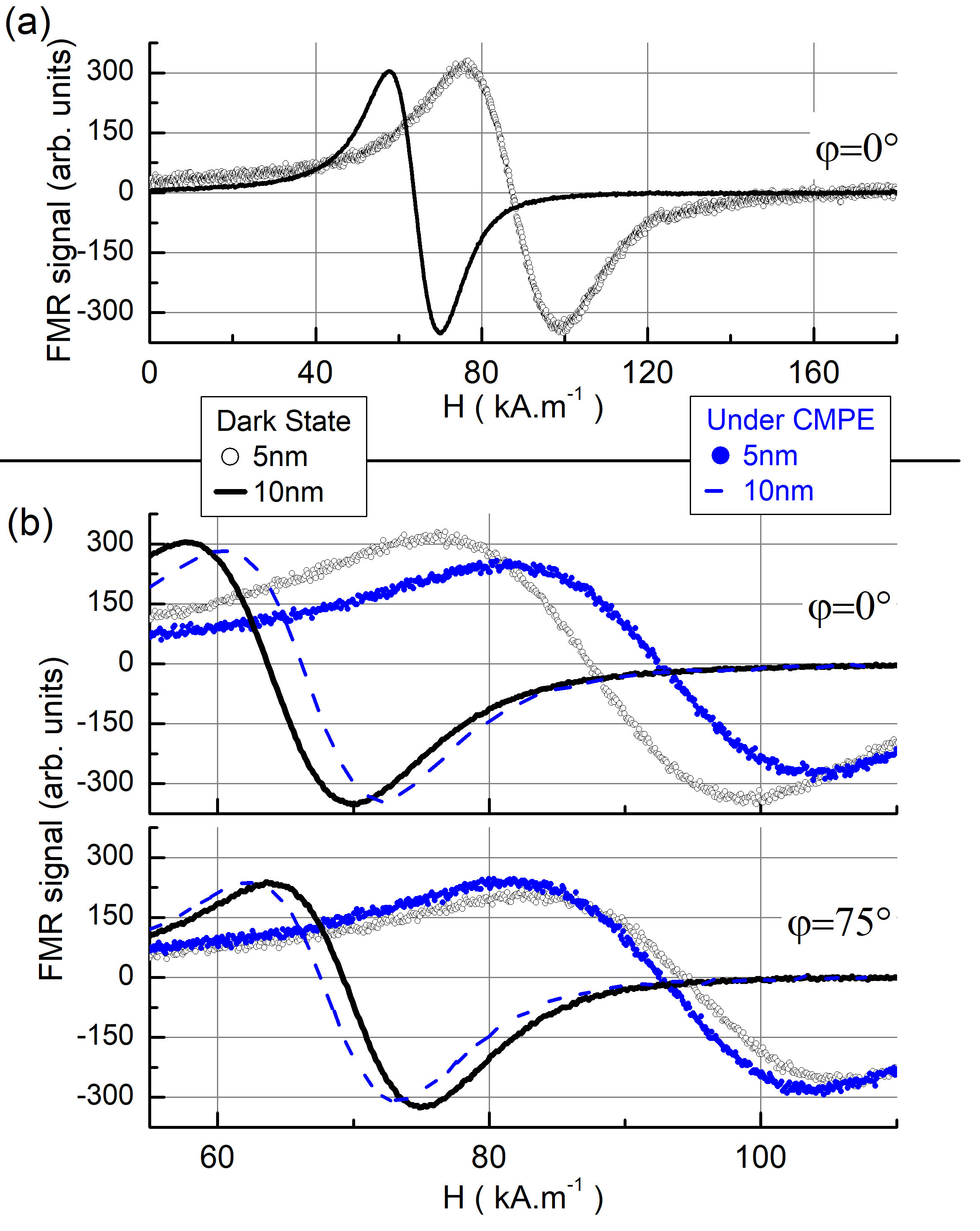}
\caption{\label{fig:fig4} (a) In-plane field-sweep FMR spectra
for the 5~nm and 10~nm thick samples measured at 9.3 GHz along $\varphi=0\degree$, in the dark state. (b) In-plane field-sweep FMR spectra
for both samples measured at 9.3 GHz under CMPE induced by the LED illumination and in the dark state, along $\varphi=0\degree$ (top) and along $\varphi=75\degree$ (bottom).}
\end{figure}

Using Ferromagnetic Resonance (FMR), we probed the dynamic control of magnetization on the 5~nm and 10~nm FeGa thin films illuminated with an intensity of 7~W.cm$^{-2}$ from a 405~nm LED. It should be noted here that the LED could not be used for static measurements due to space restrictions within the MOKE apparatus. Figure~\ref{fig:fig4} (a) shows Ferromagnetic resonance (FMR) spectra obtained in the dark state and measured with the dc external magnetic field applied along $\varphi=0\degree$ (while the magnetic component of the microwave field was perpendicular to the dc field). It reveals a FMR lineshape with the resonance field (the external magnetic field at which the power absorption spectrum $dI/dH$ crosses zero, i.e. maximum power absorption) at 87.5 kA.m$^{-1}$ (63.7 kA.m$^{-1}$) for the 5~nm (10~nm) thick FeGa.



For both thicknesses, no significant modifications of the FMR spectra were observed under 0.6~W.cm$^{-2}$ illumination with the 410~nm laser diode. However, significant modifications of the FMR signal were obtained under CMPE with the 7~W.cm$^{-2}$ LED illumination at 405~nm as shown in Figure~\ref{fig:fig4} (b). The FMR resonance field increases significantly under CMPE, along $\varphi=0\degree$. On the opposite, the FMR resonance field decreases under CMPE along $\varphi=75\degree$.\ 

\begin{figure}
\centering\includegraphics[width=11cm]{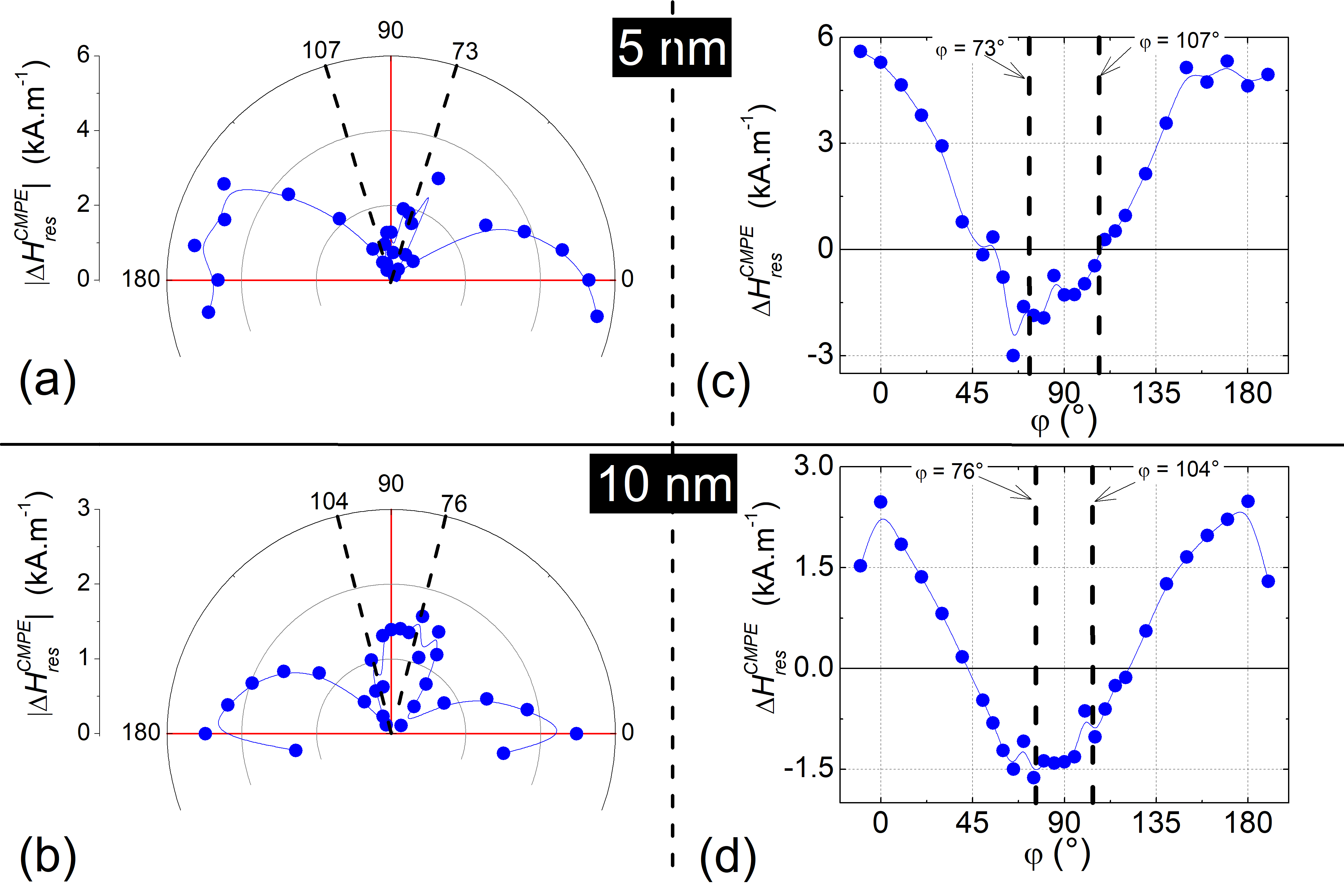}
\caption{\label{fig:fig5} Angular dependencies of $ \lvert \Delta H^{CMPE}_{res} \rvert $ for the 5~nm thick sample (a) and for the 10~nm thick sample (b). Angular dependencies of $\Delta H^{CMPE}_{res}$ for the 5~nm thick sample (c) and for the 10~nm thick sample (d). (Dashed lines indicate the hard axis angular positions, in the dark state, as shown in Figure 1, 2 and 3.)}
\end{figure}

To further understand the angular dependence of FeGa dynamic properties modifications under CMPE, systematic FMR angular dependent studies were performed. For both thicknesses, angular dependencies of resonant field shifts under CMPE ($\Delta H^{CMPE}_{res}$) exhibit extrema along global easy and hard axes as shown in Figure~\ref{fig:fig5}. Extrema are positive in the vicinity of the global easy axis but negative in the vicinity of the global hard axis as shown in Figure~\ref{fig:fig5} (c) and (d). It shows that significant resonant field shifts are achieved and they are prominent in the vicinity of their anisotropy axes.
Resonant field shifts are known to occur in ExMF under electric field~\cite{Liu_Philosophical_2014}. As a matter of comparison with our results, electrical field induced positive and negative shifts of the resonance field as a function of the in-plane azimutal angle were previously observed in FeGa/PMN-PT heterostructures~\cite{Jimenez_JMMM_2020}. These shifts were shown to arise from electrical field induced modifications of FeGa anisotropy. In the study presented here, resonant field shifts of different signs are shown to be induced not by an electrical field but by an illumination.

Under CMPE with the LED illumination, the maximum value of $\Delta H^{CMPE}_{res}$ is +5.0~kA.m$^{-1}$ (2.5~kA.m$^{-1}$) for the 5~nm (10~nm) thick sample. It represents a 5.7\% (4.0\%) increase of the resonant field under CMPE for the thinner (thicker) sample. 
It confirms a CMPE decrease with increasing FeGa thickness, as determined by the static measurements. FeGa FMR linewidths, which is a critical parameter for microwave magnetic materials, are not modified by the CMPE.\


It should be noted here that the temperature increase due to the LED illumination during FMR measurements was measured and found to be under 6~K (see Figure S5 of the supplemental information). In principle, the magnetization of a ferromagnet decreases because of the heating and this thermal effect results in a positive shift of a resonance field. Here, the presence of a negative $\Delta H^{CMPE}_{res}$ indicates that the driving mechanism for this shift is not related to the magnetization decrease due to heating. Furthermore, in the dark state, a study of the thermal dependence of the resonance field position revealed a shift of +0.04 kA.m$^{-1}$.K$^{-1}$ along $\varphi=0\degree$. Given that thermal dependence, the maximum induced resonant field shift due to a 6K heating is calculated to be +0.24 kA.m$^{-1}$, which is very little as compared to resonant field shifts under illumination observed in our study. Also, resonant field shifts and temperature changes do not exhibit similar time dependence (see Figure S5 of the supplemental information). The observed light-induced shift is thus mainly attributed to the CMPE and results presented here are experimental evidences for the control of dynamic properties by CMPE.


\newpage
\section{Discussion}

In summary, our study reveals that not only that static but also dynamic magnetic properties of an extrinsic multiferroic can be modified by CMPE under visible light. About static properties, magnetization reversals of 5~nm and 10~nm FeGa thin films on PMN-PZT substrates are shown to be modified by CMPE. The angular dependent study shows that M-H loops area, coercive fields and remanent magnetizations are enhanced or reduced under CMPE according to the direction of the applied field.
A converse magneto-photostrictive coefficient, $\alpha_{CMP}^\lambda$, is defined in this manuscript. It provides a general approach to analyse and to compare different ExMF under CMPE. $\alpha_{CMP}^{410~nm}$ angular-dependent shapes and magnitudes are thickness-dependent but its magnitude always presents maxima in the vicinity of global easy axes. When compared, the angular dependence for the $\alpha_{CMP}^{410nm}$ and for the converse magneto-electric coefficient are similar in shape.\ 

 We reveal that dynamic properties of magnetization probed by FMR are shown to be significantly modified under CMPE with visible light. Resonance fields are shifted under CMPE whereas their linewidths remain constant. Furthermore, these shifts are either positive or negative as a function of the in-plane angle, with maxima along global easy axes. The largest shift of 5.7\% is obtained for the thinnest sample along the global easy axis.\ 

Our study demonstrates that the CMPE constitutes an alternative and complementary approach to the use of electric fields for a control of static and dynamic magnetic properties in multiferroics. In general, the work presented here provides an experimental foundation for a better understanding of ExMF materials and properties. The CMPE dependence on the illumination characteristics (intensity, wavelength, polarization) should be of interest for a further understanding of the light-matter interaction encountered here, for the FeGa/PMN-PZT system and for other ExMF. Furthermore, to reduce slow response times observed in our study, utilizing the faster microscale photovoltaic response and construction of local photostrictive strain should also be of interest.\ 

In general, the work presented here provides a path towards understanding the light-induced magnetization changes. It shines a new light on use of photostriction to control not only static but also dynamic magnetic properties. These remote controls present a potential to be used in wireless and energy efficient approaches to control magnetic properties and tunable RF/microwave devices.

\newpage

\section{Supplementary Information}

Scheme of the experimental set-up, entire M-H loops of the normalized magnetization reversal, time evolution of temperature changes, time evolution under sample illumination of the M-H loops, time evolution of the temperature and of the resonant field, can be found in the supplementary information.

\section{Acknowledgements}
Authors acknowledge the financial support from the South African National Research Foundation (Grant No: 88080) and the URC/FRC, University of Johannesburg (UJ), South Africa. Authors wish to acknowledge the support of University of Brest in funding ML’s Ph.D, the support of Brest Metropole in funding G.B's Ph. D, the support of the Agence Nationale de la Recherche in funding L.T's Ph. D, and the support of the IBSAM institute in funding the collaborative MINOTAUR project. Authors wish to acknowledge the technical support of G. Mignot for the MOKE experimental set-up.

\section{Methods}

\textbf{Sample preparation} : the samples were prepared by depositing the magnetostrictive FeGa thin films onto the PMN-PZT substrates, using radio-frequency magnetron sputtering. PMN-PZT rhombohedral single crystals grown by solid-state crystal growth are commercially available as “CPSC160-95” from Ceracomp Co. Ltd., Korea~\cite{ceracomp}. The PMN-PZT substrates used in this study had physical dimensions of 0.3~mm thick, 3~mm wide and 5~mm long. Initially, the PMN-PZT substrates are cleaned with ethanol and
acetone. A Fe$_{81}$Ga$_{19}$ polycrystalline target with a diameter of 3 inches is used in an Oerlikon Leybold Univex
350 sputtering system. The base pressure prior to the film
deposition is typically 10$^{-7}$ mbar. Fe$_{81}$Ga$_{19}$ thin films are deposited onto the PMN-PZT beams at room temperature using 100 W of deposition power and about 10 sccm argon flow rate. The stack was capped in situ with a 10 nm-thick Ta layer to protect the FeGa layer against oxidation. The growth was carried out under an in-plane magnetic field $H$\textsubscript{dep} = 2.4 kA.m$^{-1}$ along the [100] direction of the PMN-PZT substrate. Further information on the poling procedure, growth conditions, growth preparations can be found in previous publications \cite{jay2019, Jahjah2020_PRA}.\

\textbf{Static magnetic measurements} : magnetic measurements at room temperature were determined using the magneto-optic Kerr effect (MOKE) in a wide-field Kerr microscope from \textit{Evico Magnetics}~\cite{Evico}. In order to improve the signal-to-noise ratio, any MRev loop presented in this manuscript results from the average of 5 MRev loops acquired during 150~s. The CMPE control of static magnetic properties was studied by illuminating the samples with an intensity $I_1$ of 0.6~W.cm$^{-2}$ from a 410~nm laser diode. The laser spot size is (570$\times$1776) $\mu$m$^{2}$ (FWHM) and the incident angle of the laser beam on the sample surface is 20$\degree$. The samples were illuminated on the backside (i.e. the substrate side). For further information, a scheme of the experimental setup can be found in the supplementary information (see Figure S1). When the substrate was illuminated, the magnetic measurements were offset by a delay time of 60~s. Increasing the delay time did not change the results of the static magnetic measurements presented in this manuscript. The temperature was probed during measurements using an infrared pyrometer.\

\textbf{Dynamic magnetic measurements}: an \textit{Elesxys 500 Bruker} electron paramagnetic resonance spectrometer operating at X-band (9.3 GHz) was used to characterize the microwave performance of FeGa /PMN-PZT heterostructures. For dynamic measurements, the CMPE was studied by illuminating the samples with the laser diode (the one previously described in the Static magnetic measurements) and it was also studied by illuminating the samples with a LED. The LED wavelength is 405~nm and the intensity of the LED illumination is 7 W.cm$^{-2}$. The incident angle of the beam on the sample surface is 0$\degree$ (i.e. along the surface normal). The samples were illuminated on the backside (i.e. the substrate side). When the substrate was illuminated, the magnetic measurements were offset by a delay time of 360~s. Increasing the delay time did not change significantly the results of the dynamic magnetic measurements presented in this manuscript, as shown in the Figure S5 of the supplementary information.

\section{Authors contributions}

D.T.D, A.F, J-Ph.J, D.S conceived and directed this project. D.T.D, A.F, J-Ph.J, M.L, D.S realized the samples. D.T.D, M.D, A.F, W.J, J-Ph.J, Y.LG, M.L, A.R.E.P, C.S, D.S performed and analyzed static measurements. G.B, V.C, D.T.D, A.F, J-Ph.J, M.L, G.S, D.S, L.T-K, V.V performed and analyzed dynamic measurements. This manuscript was mainly prepared by D.T.D, A.F, J-Ph.J, M.L, D.S and all authors contributed to the discussion of the results and the text.

\newpage


\providecommand{\latin}[1]{#1}
\makeatletter
\providecommand{\doi}
  {\begingroup\let\do\@makeother\dospecials
  \catcode`\{=1 \catcode`\}=2 \doi@aux}
\providecommand{\doi@aux}[1]{\endgroup\texttt{#1}}
\makeatother
\providecommand*\mcitethebibliography{\thebibliography}
\csname @ifundefined\endcsname{endmcitethebibliography}
  {\let\endmcitethebibliography\endthebibliography}{}

\end{document}



\title{Supplementary information for "Converse magneto-photostrictive effect and its coefficient in extrinsic multiferroics."}


\author{Matthieu Liparo}
     \affiliation{Univ. Brest, Laboratoire d'Optique et de Magnétisme (OPTIMAG), EA 938, 29200 Brest, France}
    \affiliation{Cr Research Group, Department of Physics, University of Johannesburg, PO Box 524, Auckland Park 2006, South Africa}
 
   \author{Jean-Philippe Jay}
    \author{Matthieu Dubreuil}
      \affiliation{Univ. Brest, Laboratoire d'Optique et de Magnétisme (OPTIMAG), EA 938, 29200 Brest, France}
    \author{Ga\"elle Simon}
     \affiliation{Univ. Brest, Service général des plateformes technologiques, Service RMN-RPE, 29200 Brest, France}
    \author{Alain Fessant}
    \affiliation{Univ. Brest, Laboratoire d'Optique et de Magnétisme (OPTIMAG), EA 938, 29200 Brest, France}
\author{Walaa Jahjah}
\author{Yann Le Grand}
     \affiliation{Univ. Brest, Laboratoire d'Optique et de Magnétisme (OPTIMAG), EA 938, 29200 Brest, France}
\author{Charles Sheppard}
\author{Aletta R. E. Prinsloo}
    \affiliation{Cr Research Group, Department of Physics, University of Johannesburg, PO Box 524, Auckland Park 2006, South Africa}
    \author{Vincent Vlaminck}
    \author{Vincent Castel}
    \author{Loic Temdie-Kom}
    \author{Guillaume Bourcin}
\affiliation{IMT-Atlantique, Campus de Brest, Département Micro-Ondes, Technopole Brest-Iroise, CS83818, 29238 Brest Cedex 03, France}
      \author{David Spenato}
   \affiliation{Univ. Brest, Laboratoire d'Optique et de Magnétisme (OPTIMAG), EA 938, 29200 Brest, France}
\author{David T. Dekadjevi}

      \affiliation{Univ. Brest, Laboratoire d'Optique et de Magnétisme (OPTIMAG), EA 938, 29200 Brest, France}
    \affiliation{Cr Research Group, Department of Physics, University of Johannesburg, PO Box 524, Auckland Park 2006, South Africa}
    \email{david.dekadjevi@univ-brest.fr}


\date{\today}



\maketitle

\newpage

Figure S1 shows a scheme of the experimental set-up for static measurements. \\

\begin{figure}[h!]
\centering\includegraphics[width=15cm]{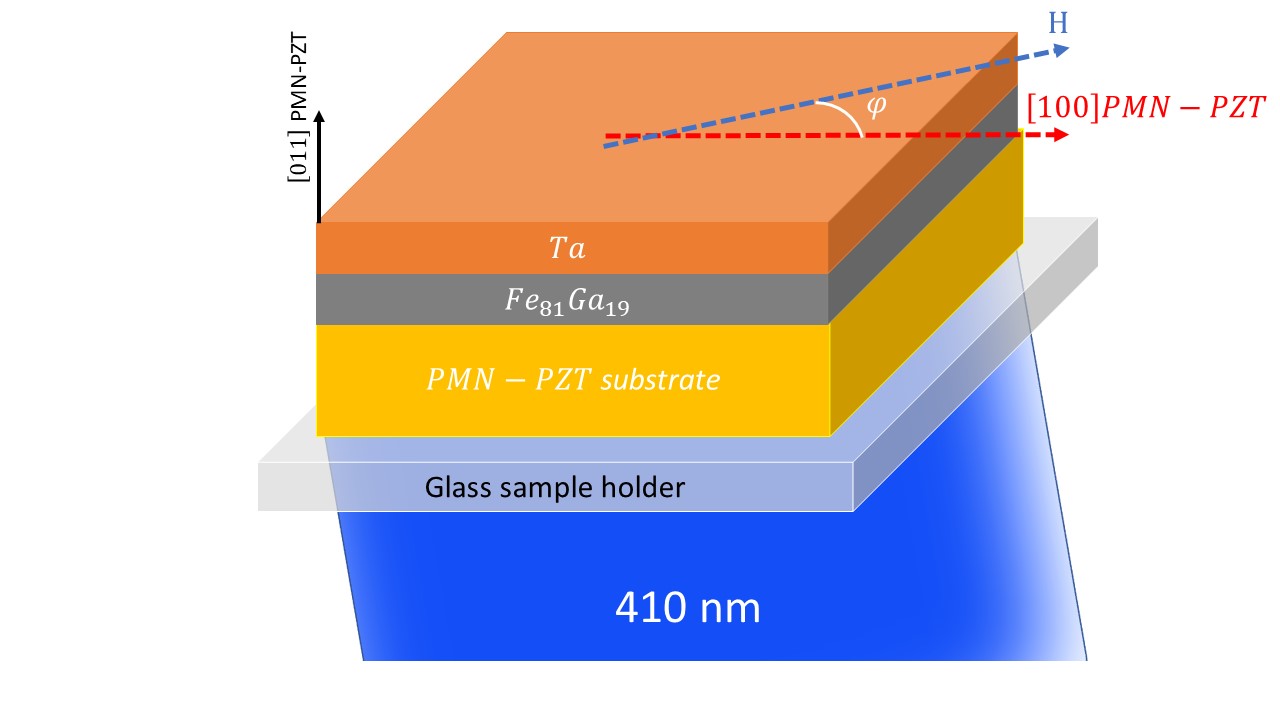}
\caption{\label{fig:fig1} Schematic drawing of the ME composite consisting of the magnetostrictive FeGa and the (011)-oriented PMN-PZT substrate. The configuration of measurement is also indicated using $\varphi$, the angle between the applied magnetic field $H$ (dashed blue axis) and the [100] direction of the substrate (dashed red axis).}
\end{figure} 

\newpage

Figure S2 (a) and (b) show the entire M-H loops which were zoomed to obtain Figure 1(a) and (b) of the manuscript.

\begin{figure}[h!]
\centering\includegraphics[width=12cm]{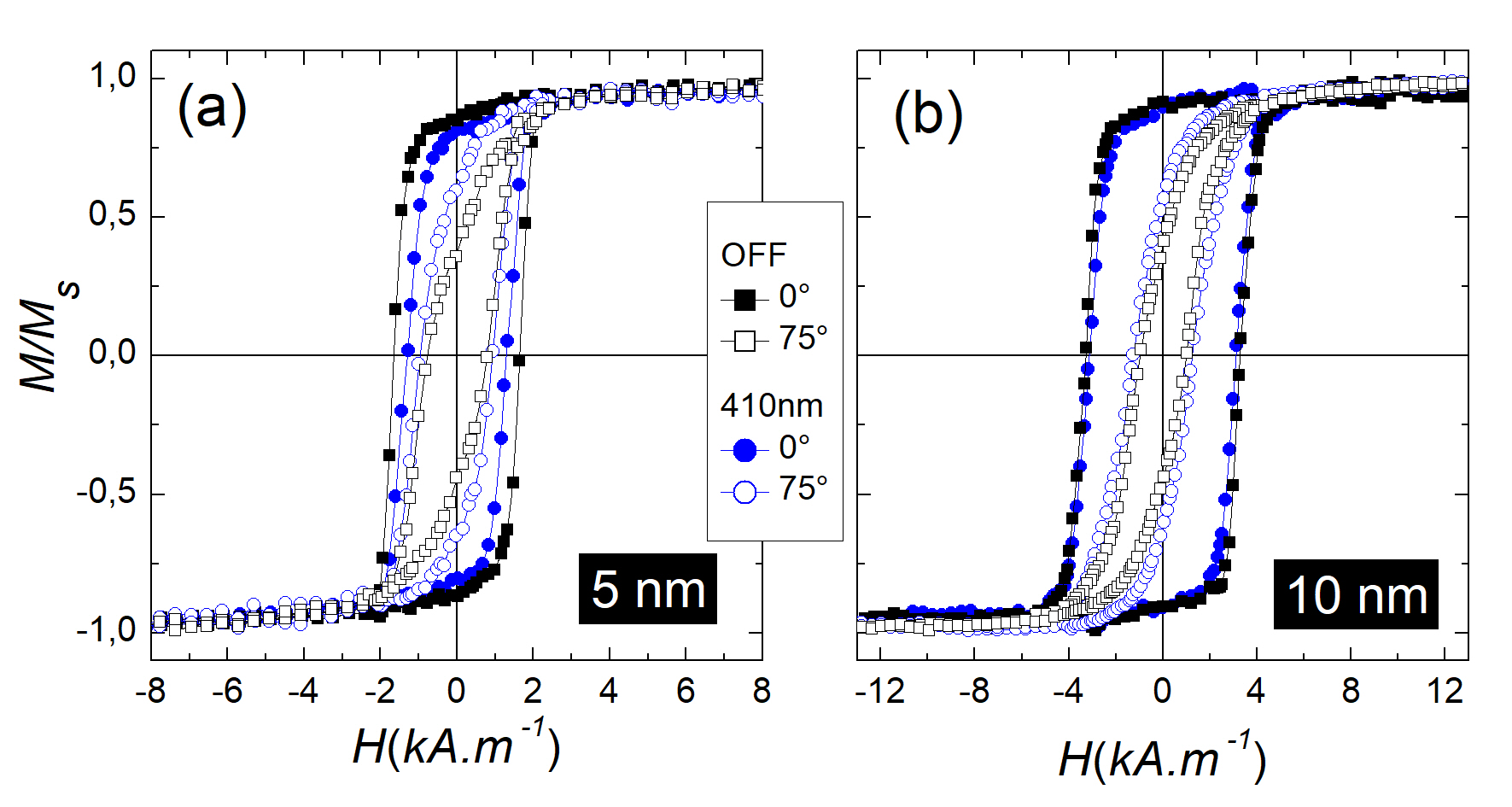}
\caption{\label{fig:fig1a} The entire MH loops of the normalized magnetization reversal of the Ta(10 nm)/FeGa(5 nm ; 10 nm)/PMN-PZT(0.3 mm), measured in-plane with the magnetic field $H$ respectively parallel to [100] ($\varphi$ = 0\degree, i.e. along the deposition field axis $H$\textsubscript{dep}) and $\varphi$ = 75\degree, under MPE (i.e. under illumination at 410nm) and in the dark state (OFF) materials.}
\end{figure}

\newpage

Figure S3 shows the time evolution of M-H loops under sample illumination for the 10~nm FeGa sample at $\varphi=0\degree$. The illumination started at $t=$~0~s and a delay time of 45~s was set before starting the first M-H loop measurement. Each M-H loop took 150~s to be obtained. (In the manuscript, a given MH loop corresponds to the average of 5 MH loops. Thus, it took $5 \times 150$~s$=750$~s to record each M-H loop presented in the manuscript).\\

\begin{figure}[h!]
\centering\includegraphics[width=12cm]{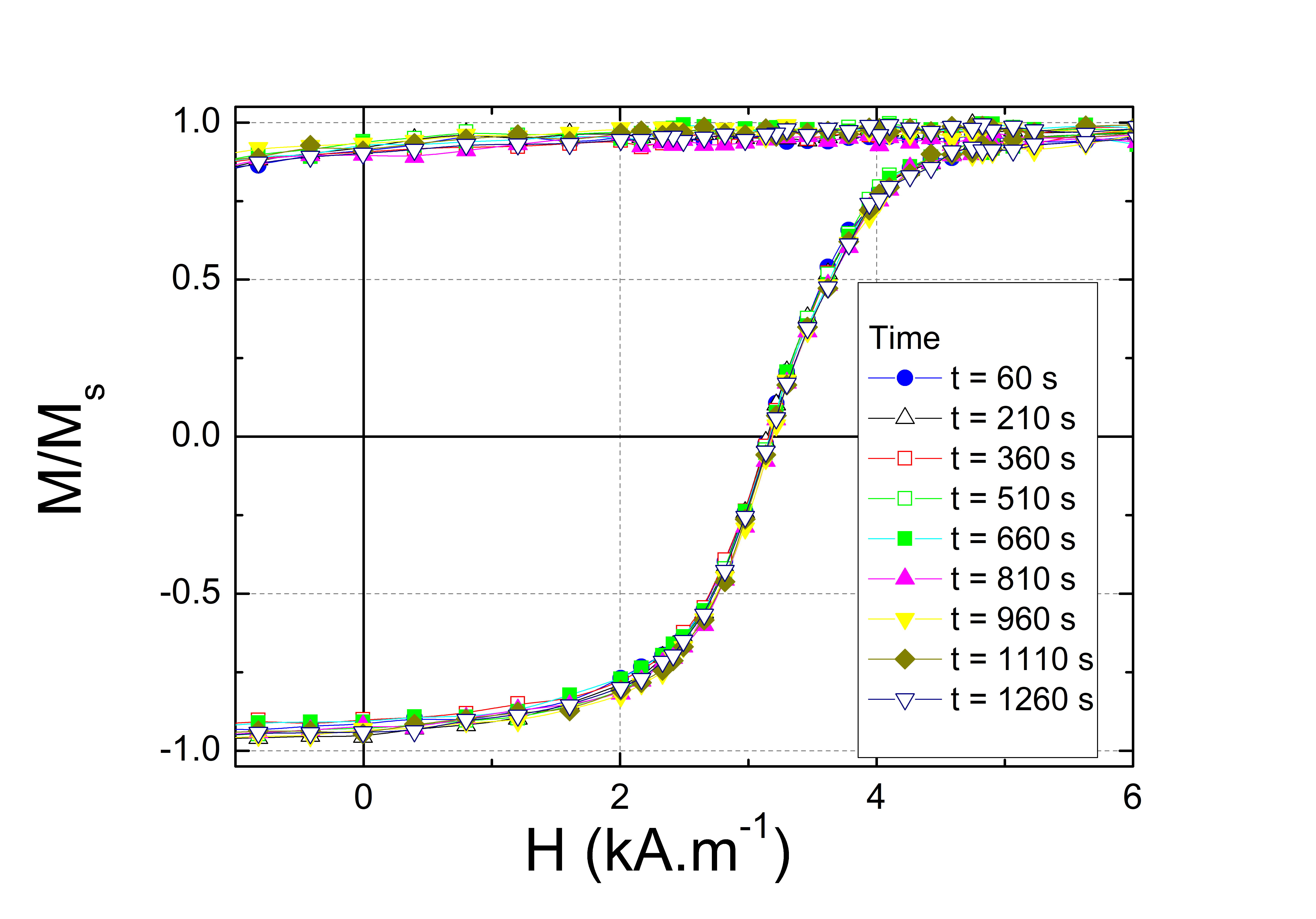}
\caption{\label{fig:fig3} Time evolution under sample illumination of the M-H loops for the 10~nm FeGa sample at $\varphi=0\degree$, zoomed in-between $H=-1$ kA.m$^{-1}$ and 6 kA.m$^{-1}$}.
\end{figure}

\newpage


Figure S4 shows the time evolution of temperature changes probed at the sample surface when the material is illuminated by the laser. The illumination started at $t=0$ s. The measurement was performed with an infrared thermometer.\\

\begin{figure}[h!]
\centering\includegraphics[width=12cm]{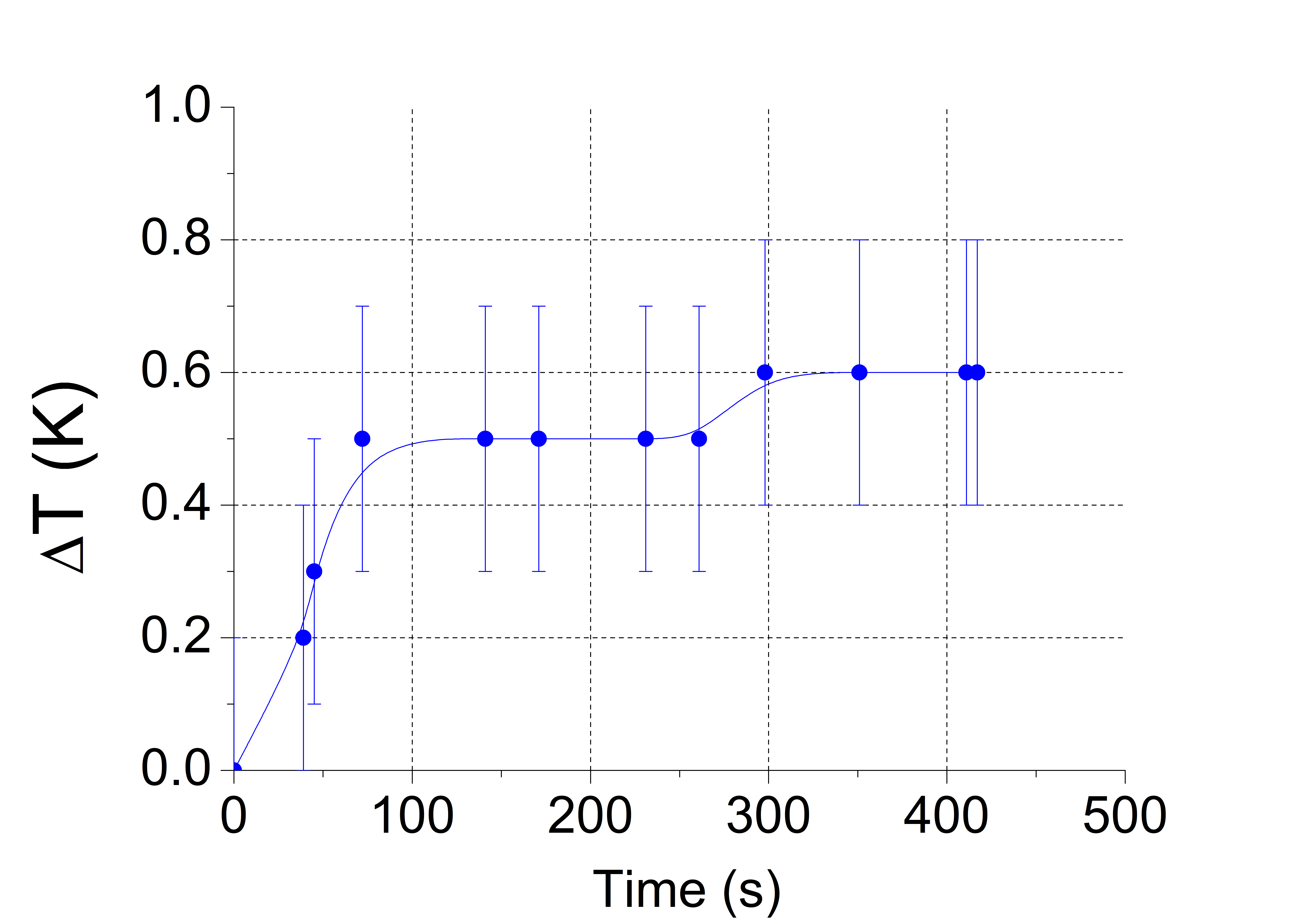}
\caption{\label{fig:fig2} Time evolution of temperature changes probed at the sample surface under laser illumination.}
\end{figure}

\newpage

Figure S5 shows the time evolution of the temperature and of the resonance field for the 5~nm sample along $\varphi=0\degree$. The LED illumination began at t=125s (i.e. LED ON) and finished at t=1836s (i.e. LED OFF).

\begin{figure}[h!]
\centering\includegraphics[width=12cm]{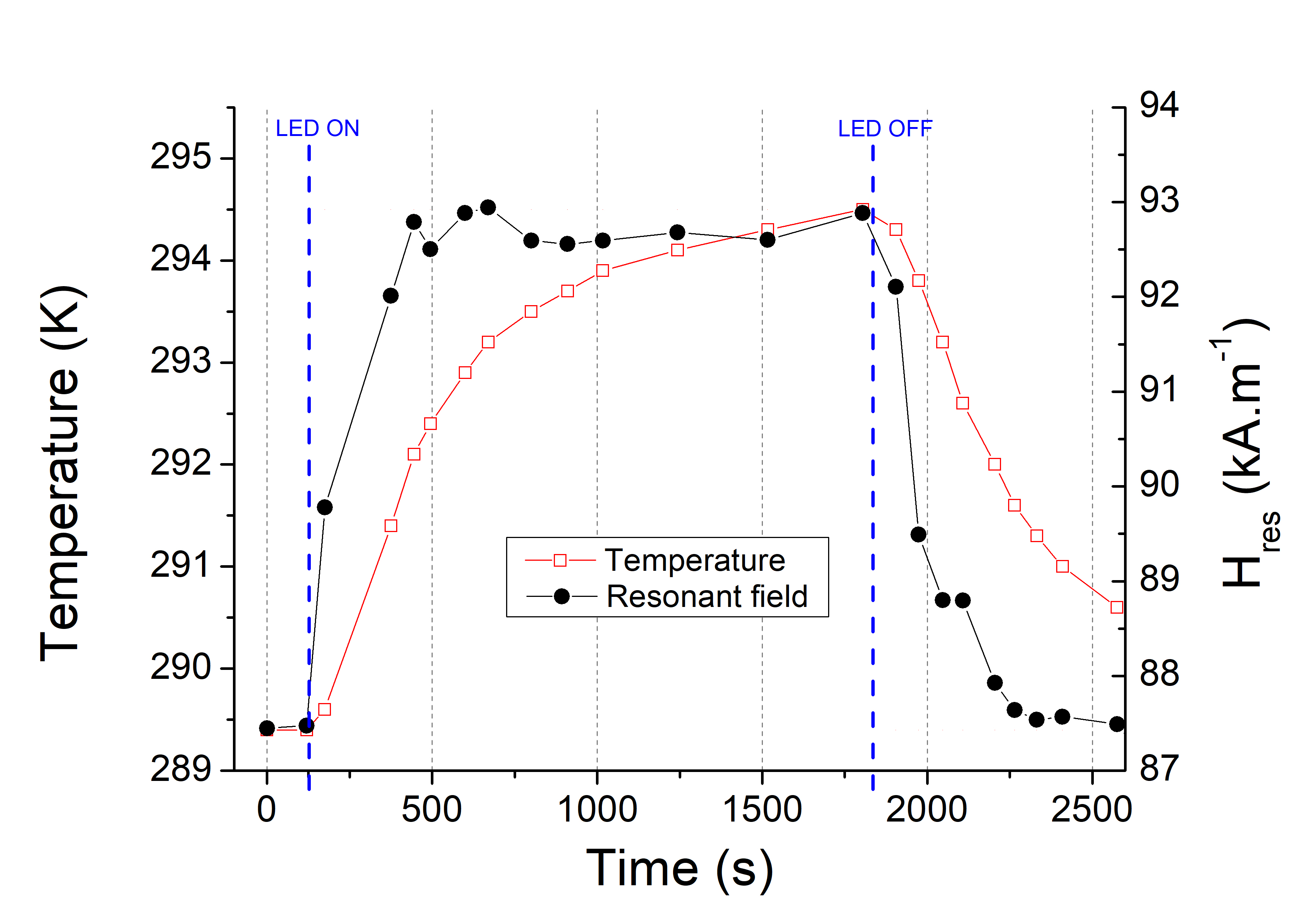}
\caption{\label{fig:fig5} Time evolution of the temperature and of the resonance field for the 5~nm sample along $\varphi=0\degree$.}.
\end{figure}